\documentclass[12pt]{article}

\usepackage{graphicx}

\graphicspath{C:/Users/hs002673/Desktop/clustering/}

\usepackage{amsmath}

\usepackage{amssymb}

\usepackage{amsmath,bm}

\usepackage{epsfig}

\usepackage[authoryear]{natbib}

\usepackage{algorithm}

\usepackage{algpseudocode}% http://ctan.org/pkg/algorithmicx

\newtheorem{thm}{Theorem}

\newtheorem{lem}{Lemma}

\newcommand{\Real}{\mathbb{R}}
\newcommand{\G}{\cal{G}}
\newcommand{\GA}{\mathop{\it GA}}
\newcommand{\GY}{\mathcal{G}Y}

\begin{document}

\baselineskip 18pt

\title{An Affine-Transformation Invariant Bayesian Cluster Process with Split-Merge Gibbs Sampler}

\author{Hsin-Hsiung Huang, Jie Yang}

\maketitle

\begin{abstract}
In order to identify clusters of objects
 with features transformed by unknown affine transformations,
we develop a Bayesian cluster process which is invariant with respect to
certain linear transformations of the feature space
and able to cluster data without knowing the number of clusters in advance.
Specifically, our proposed method can identify clusters invariant to
orthogonal transformations under model I,
invariant to scaling-coordinate orthogonal transformations under model II,
 or invariant to arbitrary non-singular linear transformations under model III.
The proposed split-merge algorithm leads to an irreducible and aperiodic Markov chain,
which is also efficient at identifying clusters reasonably well for various applications.
We illustrate the applications of our approach to both synthetic and real data
such as leukemia gene expression data for model I;
wine data and two half-moons benchmark data for model II;
three-dimensional Denmark road geographic coordinate system data and
an arbitrary non-singular transformed two half-moons data for model III.
These examples show that the proposed method could be widely applied in many fields,
especially for finding
the number of clusters and identifying clusters of samples of interest
in aerial photography and genomic data.
\end{abstract}

\noindent


\section{Introduction}
Clustering of objects invariant
with respect to affine transformations of feature
vectors is an important research topic
 since objects may be recorded via different angles and positions
 so that their coordinates may vary
 and their nearest neighbors may belong to other clusters.
 For example, the longitude, latitude, and altitude coordinates
  of an object which are recorded by devices equipped in
  aircrafts or satellites change across different observation time.
In this situation, distance-based clustering method including
 $K$-means \citep{MacQueen1967}, hierarchical clustering \citep{Ward1963},
 clustering based on principal components, spectral clustering \citep{Ng2001},
  and others \citep{JainDubes1988, Ozawa1985}
  may fail to identify the correct clusters by grouping nearest points.
 Another category is distribution-based
clustering methods \citep{Banfield1993, FraleyRaftery1998, FraleyRaftery2002, FraleyRaftery2007, McCYang2008, Vogt2010}
 which may specify a partition as a parameter in a likelihood function and
estimate it under a Bayesian framework.
These existing methods typically assume
that the covariance structure is proportional to an identity matrix,
and thus may not work on general cases
in which data are distorted by an affine transformation.

In certain areas of application, the goal is to cluster objects
$i=1,\ldots, n$
into disjoint subsets based on their feature vectors $Y_i \in \Real^d$.
This paper considers three closely related cluster process that are
invariant with
respect to three groups of linear transformations $g\colon\Real^d\to\Real^d$
acting on the feature space.  Group invariance implies that the feature
configurations $Y$ and
$Y'$ in $\Real^{n\times d}$ determine the same clustering, or probability
distribution on clusterings,
if they belong to the same group orbit.  For example, if the feature space
is Euclidean and $\G$~is
the group of Euclidean isometries or congruences, the clustering is a
function only of
the maximal invariant, which is the array of Euclidean distances $D_{ij} =
\|Y_i - Y_j\|$. For example, image data such as the aerial photography
and three-dimensional protein structures are two motivating examples.
The shape and relative locations of data may vary due to the change
of the viewer's angle and positions.

\cite{McC2008} modeled the data $Y=\{Y_{i,j}\}$ as $d$ series of a stationary autoregressive
Gaussian process with mean zero, three between-series variance structures,
 and an autocorrelation function $\Gamma.$
Then the profile likelihoods of covariance and partition were derived
under three types of covariance structures
which could be (1) proportional to an identity matrix, $\sigma^2 I_d,$ (2) proportional to a diagonal matrix,
and (3) an arbitrary positive definite matrix.
These three covariance structures correspond to three kinds of
affine transformation: (1) index permutations, rotation, one-scaling on all variables,
and location-translation transformations
which are under the first type of covariance structures that is named model I and
the transformation and covariance structure $\sigma^2 I_d$ were also adopted by \cite{Vogt2010};
(2) each variable may have different scaling transformations
which are under the second type of covariance structures that is named model II;
(3) the variables are transformed by a nonsingular matrix
that is named model III, where the observed variables may be
linear combinations of
some latent variables in model I.
These models cover fairly general situations of clustering in nature.

In the literature, the use of a Dirichlet process prior prevents users from
assuming the the number of clusters before finding the partition.
In this paper, we follow \cite{McCYang2008} and
assume that
the prior on partitions of objects follows the Ewens distribution
\citep{Ewens1972}.
We also propose an efficient split-merge sampling
algorithm in generating
partition candidates while keeping
the resulting partition-valued Markov chain ergodic.

\section{Cluster process and priors}
In this paper, an $\Real^d$-valued  {\it cluster process\/} $(Y, B)$
means a random partition $B$ of the natural numbers,
together with an infinite sequence $Y_1, Y_2,\ldots$ of random vectors
in the state space~$\Real^d$.
The restriction of such a process to a finite sample
$[n]=\{1,\ldots, n\}$ of units or specimens consists
of the restricted partition $B[n]$ in company with the finite sequence
$Y[n] = (Y_1,\ldots, Y_n)$.
A partition $B[n]:[n]\times[n]\to\{0,1\}$~is the partition of the sample units
expressed as a binary cluster-factor matrix of $B_{i, j} = 1$ if $Y_i$ and $Y_j$
are of the same cluster (denoted as $i\sim j$), and $B_{i, j} = 0$ otherwise.
The term {\it cluster process\/} implies infinite exchangeability,
which means that the joint distribution $p_n$ of $(Y[n], B[n])$ is symmetric \citep{McCYang2006} or
invariant under permutations of indices
\citep{Pitman2006}, and $p_n$ is the marginal distribution of $p_{n+1}$ under deletion
of the $(n+1)$th unit from the sample.

The simplest example of such processes is the
exchangeable Gaussian mixtures constructed as follows \citep{McCYang2008}.
First, $B\sim p$ is some infinitely exchangeable random partition.
Second, the conditional distribution of the samples $Y$, which is regarded as
a matrix of order $n\times d$
given $B$ and $\theta$, is Gaussian with mean and variance as follows
$$
E(Y_{i, r} \,|\, B) = \mu_r,\qquad
{\mathop{\rm Cov}\nolimits}(Y_{i,r}, Y_{j,s} \,|\, B,\theta) = (\delta_{i,j} + \theta B_{i,j}) \Sigma_{r,s}.
%\eqno(1)
$$
where $\delta$ is Kronecker's delta, that is, $\delta_{i,j}=1$ if $i=j$ and $0$ if $i\neq j$,
$\theta$ is a positive parameter, and $\Sigma=\left(\Sigma_{r,s}\right)$ is
a positive definite matrix of order $d\times d$ for $1\leq i,j\leq n$ and $1\leq r,s\leq d.$
The mean and covariance of $Y$ given $B$ are
$$
E(Y \,|\, B) = {\bf 1}_n \bm{ \mu}^\intercal,\qquad
{\mathop{\rm Cov}\nolimits}(Y \,|\, B,\theta) = (I_n + \theta B[n]) \otimes \Sigma
$$
where $\bm{ \mu}^\intercal$ is the transpose of the feature mean vector $\bm{ \mu}=(\mu_1,\ldots,\mu_d)^\intercal$,
${\bf 1}_n$ is the vector in $\Real^n$ whose components are all one,
 and ``$\otimes$'' indicates the Kronecker product.
The identity matrix itself is also a partition in which each cluster consists of one element.

Given the number of clusters $k$,
the cluster size $(n_1,\ldots,n_k)$ follows a multinomial distribution $\bm{ \pi}=(\pi_1,\ldots,\pi_k),$
and $\bm{ \pi}$ is a random vector
from the exchangeable Dirichlet distribution Dir$(\lambda/k,\ldots,\lambda/k)$.
After integrating out $\pi$, the partition follows
 a Dirichlet-Multinomial prior
$$
p_n(B|n, \lambda, k) =\frac{ k!}{(k - \#B)!}\frac{\Gamma(\lambda)\prod_{b\in B}\Gamma(n_b+\lambda/k)}{\Gamma(n+\lambda)[\Gamma(\lambda/k)]^{\#B}},
$$
where $\#B\leq k$ denotes the number of clusters present in the partition $B$ and $n_b$ is the size of cluster $b$ \citep{MacEachern1994, Dahl2005, McCYang2008}.
The limit as $k \to\infty$ is well defined and known as the Ewens process with a distribution
 $$
 p_n(B|n,\lambda) = \frac{\Gamma(\lambda)\lambda^{\# B}}{\Gamma(n+\lambda)}\prod_{b\in B}\Gamma(n_b),
 $$
which is also known as Chinese Restaurant process \citep{Ewens1972, Neal2000, BlJordan2006, Crane2016}.

In this paper, we adopt the Ewens prior for partition $B$
which implies $k=\infty$ in the population.
Note that $\#B \leq n$ for any given sample size $n$.
\cite{McCYang2008}
provided a framework with a finite number of clusters and more general covariance structures.

We choose a proper prior distribution for the
variance ratio $\theta$, the symmetric $F$-family
$$
p(\theta)\propto \frac{\theta^{\alpha-1}}{(1+\theta)^{2\alpha}}
$$
with $\alpha > 0$ allowing a range of reasonable choices \citep{Chaloner1987}.

We propose a Gibbs sampling procedure to estimate the partition $B$
and the parameter $\theta$ from conditional probabilities.
Since the conditional distribution
of $\theta$ does not have a recognized form,
we propose to use a discrete version
$\{p(\theta_j)\}^J_{j=1},$ where $J$ is a moderately large number.

\section{Affine-transformation invariant clustering}

The conditional distribution on partitions of $[n] = \{1, \ldots, n\}$
is determined by the finite sequence $Y=(Y_1,\ldots, Y_n)$
regarded as a configuration of $n$ labeled points in $\Real^d$.
The exchangeability condition implies that any permutation
of the sequence induces a corresponding permutation in $B$,
i.e.~$p_n(B^\pi \,|\, Y^\pi) = p_n(B \,|\, Y),$ where
$Y^\pi_i = Y_{\pi(i)}$ and $B^\pi_{i,j} = B_{\pi(i), \pi(j)}$.
In many cases, it is reasonable to assume
 additional symmetries involving transformations in $\Real^d$,
 for example $p_n(B \,|\, Y) = p_n(B \,|\, -Y)$.
 We are asking, in effect, whether two labeled configurations $Y$ and $Y'$ which are
 {\it geometrically equivalent\/} in $\Real^d$ should determine the same conditional distribution on sample partitions.

If the state space $\Real^d$ is regarded as $d$-dimensional Euclidean space
with the standard Euclidean inner product and Euclidean metric,
the configurations $Y$ and $Y'$ are {\it congruent\/}
if there exists a vector $\bm{a}=(a_1,\ldots,a_d)\in\Real^d$ and an orthogonal matrix
$A\in\Real^{d\times d}$ such that $Y'_i = \bm{a} + A Y_i$ for each~$i$.
Equivalently, the $n\times n$ arrays of squared Euclidean distances
$D_{ij} = \|Y_i - Y_j\|^2$ and $D'_{ij} = \|Y'_i - Y'_j\|^2$ are equal.
The configurations are geometrically {\it similar\/}
if $Y'_i = \bm{a} + b Y_i$ for $b \in \Real$ and $b \neq 0$,
implying that the arrays of distances are proportional $D' = b^2 D$.
After respecting an observation as a group orbit,
without loss of generality we can assume that
 there is a representative element of the group orbit
 with feature mean vector $\bm{ \mu}={\bf 0}_d$,
so that $Y\sim N({\bf 0}_{n\times d}, (I_n + \theta B)\otimes b^2I_d)$.
The set of linear transformations $\{A=b I_d |b\neq 0\}$
forms a group $\Real^d \times \Real/\{0\}:\;
 y_{ij} \mapsto a_j + b y_i$ for
 $\bm{a}\in\Real^d, b\neq 0.$
 %${\mathop{\it GA}\nolimits}(\Real).$
This is model~I, which is the case considered in \cite{Vogt2010}.

In essence, the observation is not regarded as a point in $\Real^{n\times d}$
but is treated as a {\it group orbit\/} generated by the group of rigid transformations,
or similarity transformations if scalar multiples are permitted.
In statistical terms, this approach meshes with the sub-model in which the matrix $\Sigma$ in (1)
is a scaled identity matrix $I_d$.
An equivalent way of saying the same thing for $n> d$ is that the column-centered sample matrix
$\tilde Y = Y - {\bf 1}_n {\bf 1}_n^\intercal Y /n$
determines the sample covariance matrix
$S = (\tilde Y ^\intercal \tilde Y)/(n-1)$ and hence the Mahalanobis metric
$\|x - x^*\|^2 = (x - x^*)^\intercal S^{-1} (x - x^*)$
in the state space. One implication is that the $n\times n$ matrix $D = (D_{ij}) = (\|Y_i - Y_j\|^2)$ of standardized inter-point
Mahalanobis distances is maximal invariant, and the conditional distribution on sample partitions depends on $Y$
only through this matrix.

In practice, the $d$~variables are sometimes measured on scales that are not commensurate
with one another, so the state space seldom has a natural metric.
In this case, we assume that $Y$ and $Y'$
as equivalent configurations for each feature
$Y_{\cdot,j}$ if there is a vector $a_j,\; b_j \in\Real^d$
such that $Y'_{\cdot,j} = a_j + b_j Y_{\cdot,j}$.
After respecting an observation as a group orbit,
without loss of generality we can assume that
 there is a representative element of the
 group orbit with feature mean vector $\bm{ \mu}=\bm{0}_d$,
so that $Y\sim N\left({\bf 0}_{n\times d}, (I_n + \theta B)\otimes \left(\mbox{diag}(b_1^2,\ldots,b_d^2)I_d\right)\right)$.
The element of the set of linear transformations $\left\{A=\mbox{diag}(b_1,\ldots,b_d)|b_i\neq 0,\; i=1,\ldots,d\right\}$
is essentially the group
 $\GA(\Real)^d:\; y_{ij} \mapsto a_j + b_j y_{ij}$
for  $\bm{a}\in\Real^d, b_j \neq 0$
that is the general affine group acting independently on the
$d$ columns of $Y.$
No linear combinations are permitted here, so that the integrity of the
variables is preserved.
This is model II.

Moreover, in some cases, the location information or shapes of objects from aerial photography applications
may be distorted by the viewer's angle or position so that the variables may be strongly correlated.
A more extreme approach avoids the metric assumption by regarding $Y$ and $Y'$
as equivalent configurations if there exists a vector $a\in\Real^d$ and a non-singular matrix
$A\in\Real^{d\times d}$ such that $Y'_i = \bm{a} + A Y_i$
with $A^\intercal A$ is a positive definite matrix for all~$i$.
This is the general affine group $\GA(\Real^d):\; y_i \mapsto \bm{a} + A y_i$ acting
component-wise on the sequence.
For $n\le d+1$, the action is essentially transitive in the sense
that all configurations of $n$ distinct points in $\Real^d$ belong to the same orbit:
all other orbits are negligible in that they have Lebesgue measure zero.
As a result, the observation $Y$ regarded as a group orbit ${\cal G} Y$ is uninformative for clustering unless $n > d+1$.
Consequently there is a congruent group orbit with mean $\mu=0$,
$Y\sim N({\bf 0}_{n\times d}, (I_n + \theta B)\otimes A^\intercal A),$
where $A^\intercal A \in PD_d$ and $PD_d$
is the collection of $d\times d$ symmetric positive definite matrices.
This is model~III.

\subsection{Gaussian marginal probabilities}

The big advantage of regarding the observation $Y$ as a group
orbit rather than a point is that the partition of $Y$ is affine invariant and
the same as the partition of the group orbit ${\cal G}Y \subset \Real^{n\times d}$,
which is independent of the mean ${\bf 1}_n \bm{ \mu}^\intercal$.
Consequently, the distribution of the column-centered group orbit, ${\cal G}Y,$
is assumed as a Gaussian distribution
$$
N({\bf 0}_{n\times d}, (I_n + \theta B)\otimes A^\intercal A)
$$
depends only on $I_n + \theta B$ and $A^\intercal A$.
%The three groups are group I $\Real^d \times \Real^+:\;
% y_{ij} \mapsto a_j + b y_i$ for
% $\bm{a}\in\Real^d, b\neq 0,$
%group II $\GA(\Real)^d:\; y_{ij} \mapsto a_j + b_j y_{ij}$
%for  $\bm{a}\in\Real^d, b_j \neq 0,$
%and group III $\GA(\Real^d):\; y_i \mapsto \bm{a} + A y_i$.
%For group I, with $\Sigma\propto I_d$ in the Gaussian model, the likelihood
%depends only on the distance matrix $D$, so the likelihood is constant on the orbits
%associated with the larger group of Euclidean similarities.
%

\cite{McC2008} studied the $d$ series with an autocorrelation $\Gamma$
and $n$ observations in time or space following three Gaussian distribution models
$
N({\bf 0}_{n\times d}, \Gamma \otimes\Sigma)
$
under three assumptions of $\Sigma$ as follows :
\begin{eqnarray}
  \mbox{Model I: }\Sigma &=& \sigma^2 I_d, \\
  \mbox{Model II: }\Sigma &=& diag\{\sigma^2_1,\cdots,\sigma^2_d\}, \\
  \mbox{Model III: }\Sigma &\in& PD_d.
  \end{eqnarray}
These three models correspond to our three models of affine transformed group orbits which we discussed in the previous section.
In this paper, we set $(1+\theta B)$ as $\Gamma$
and $A^\intercal A$ as $\Sigma$, and then
the log likelihood based on $Y$ for all three models is obtained as follows:
\begin{eqnarray*}
  l(\Gamma,\Sigma | Y) &=& -\frac12 \log\det(\Gamma\otimes\Sigma)-\frac{1}{2} \mbox{tr}(Y'\Gamma^{-1}Y\Sigma^{-1})\\
   &=& -\frac{d}{2}\log \det(\Gamma)-\frac{n}{2}\log\det(\Sigma)-\frac{1}{2}\mbox{tr}(Y'\Gamma^{-1}Y\Sigma^{-1}),
\end{eqnarray*}
where $\Gamma^{-1}=I_n-\theta W B$, $W=\mbox{diag}(w)$, $w$ is a vector with entries $w_i=1/(1+\theta N_{ii})$, and $N=\mbox{diag}(B\bf{1}_n)$.
After plugging in the maximum likelihood estimator of $\Sigma$ which
for model III is $\hat{\Sigma}_{\Gamma}=Y'\Gamma^{-1}Y/n$,
for model II is $\mbox{diag}(\hat{\Sigma}_{\Gamma})$,
and for model I is $\mbox{tr}(\hat{\Sigma}_{\Gamma})I_d/d$, the profile likelihood of $\Gamma$, a function on orbits
(constant on each orbit), is
$$
L_p(\Gamma^{-1} | \GY) = \left\{\begin{array}{ll}
    {\mathop{\rm det}\nolimits}(\Gamma^{-1})^{d/2} / {\mathop{\rm tr}\nolimits}(Y' \Gamma^{-1} Y) ^{nd/2} & \hbox{(I)} \\
    {\mathop{\rm det}\nolimits}(\Gamma^{-1})^{d/2} / \prod_{r=1}^d(Y_r' \Gamma^{-1} Y_r)^{n/2} & (II) \\
    {\mathop{\rm det}\nolimits}(\Gamma^{-1})^{d/2} / \det(Y' \Gamma^{-1} Y)^{n/2} & (III)
\end{array}\right..
$$

The conditional distribution on partitions of $[n]$ depends on
the group orbit and the assumptions made regarding $\Sigma$.
For group I, with $\Sigma\propto I_d$ in the Gaussian model, the likelihood
depends only on the distance matrix $D$, so the likelihood is constant on the orbits
associated with the larger group of Euclidean similarities
Therefore, for model~I, the similarity transformation can be generalized as
if $Y'_i = \bm{ a }+ A Y_i$ for $A^\intercal A = \sigma^2 I_d$ and $\sigma \neq 0$,
implying that the arrays of distances are proportional $D' = \sigma^2 D$.
Consequently, there is a representative element of the group orbit
with feature mean vector $\bm{ \mu}=\bm{0}_d$,
so that $Y\sim N({\bf 0}_{n\times d}, (I_n + \theta B)\otimes \sigma^2 I_d)$.
%This is to work with ${\mathop{\it GA}\nolimits}(\Real).$

For model~II, the affine transformation can be generalized as $Y'_i = a + A Y_i,$
where $a\in\Real^d$  and a matrix
$A\in\Real^{d\times d}$
with $A^\intercal A$ as a diagonal matrix with non-zero diagonal entries for all~$i$.
As a result, there is a representative element of
the group orbit with feature mean vector ${\bm \mu}={\bm 0}_d$,
so that $Y\sim N\left({\bf 0}_{n\times d}, (I_n + \theta B)\otimes\left( \mbox{diag}(\sigma^2_1,\ldots,\sigma^2_d)I_d\right)\right)$.
This is to work with ${\mathop{\it GA}\nolimits}(\Real)^d $
which is the general affine group acting independently on the
$d$ columns of $Y$. For model~III, $\Sigma$ is an arbitrary matrix in $PD_d$.
The group is ${\mathop{\it GA}\nolimits}(\Real^d)$
and $n>d+1$. These three models are nested by
 $\rm  model \;I \subset model \;II \subset model \;III.$

Affine invariance in $\Real^d$ is a strong requirement, which comes at a small cost for moderate~$d$ provided that $d/n$ is small.
If $d/n < 1$ is not small, model III will work,
 but $Y' \Gamma^{-1} Y$ may be ill-conditioned \citep{Dempster1972, Stein1975}.
In this case, $\det(Y' \Gamma^{-1} Y)$ is close or equal to zero, so that the resulting
profile likelihood discussed becomes unstable.
In contrast, model II that is less computationally expensive than model III, and model I is the most efficient one.
%\subsection{Prior distributions of the variance ratio $\theta$ and partition $B$}

\section{Markov chain Monte Carlo sampling}

We use the prior and posterior of $\theta$ and $B$ discussed in the previous section through
a Markov chain Monte Carlo (MCMC) algorithm for estimation.
The iterative $\theta$ is obtained by Gibbs sampling \citep{Geman1984} according to the
conditional distribution
$p_n(\theta_j|B,{\cal G}Y)\propto p(\theta_j) \times L_p(\Gamma^{-1} | {\cal G}Y) ,$
where $p(\theta_j)\propto \theta_j^{\alpha-1}/(1+\theta_j)^{2\alpha}$ for $j=1,\ldots,J.$
For instance, $\alpha=1$ and the discrete set $2^{-3,\ldots,10}$ for
the range of $\theta$
are used as the default setting in our experiments.
For updating $B$, the conditional distribution on partitions is
$$
p_n(B |\theta, {\cal G} Y) \propto  p_n(B|n,\lambda) \times L_p(\Gamma^{-1} |\theta, {\cal G}Y),
$$
where $p_n(B|n,\lambda)$ is the Ewens distribution,
and a Metropolis-Hasting algorithm \citep{Hasting1970}
is used to choose the iterative $B$.
After burning in a certain number of the resulting Markov chain,
we use the average of the partition matrix as the similarity matrix to
make inference on partition.
Notice that for Algorithm 1 as follows, the transition probability $q(B^{*}|B^{(k)})=q(B^{(k)}|B^{*})$.

\begin{algorithm}[H]
\caption{MCMC algorithm}\label{Gibbs1}
\begin{algorithmic}[1]
   \State Update $B$ and $\theta$
   \For{$k =1:N$}\Comment{$N$ is the number of total iterations. Suppose that the current values are $\theta^{(k)}$ and $B^{(k)}$.}
      \State Randomly sample $\theta^{(k+1)}$ from the discrete posterior of $\theta$
      \State Randomly select an element $y_i$. Suppose $y_i$ belongs to a cluster $b_i\in B^{(k)}$
      \State Randomly assign $y_i$ into a cluster $b_j\in B^{(k)}$ other than $b_i$
      \State In the case that $|b_i|\geq 2$, $b_j$ can be an empty cluster.
      \State Call the new partition $B^*$.
      \State $R = \frac{p_n(B^*|n,\lambda)L_p(B^*| \theta^{(k+1)},{\cal G} Y)}{p_n(B^{(k)}|n,\lambda)L_p(B^{(k)}|\theta^{(k+1)},{\cal G} Y))}$
      \State Accept $B^{(k+1)}=B^{*}$ with probability $\min \{1,R\}$
      \State Keep $B^{(k+1)}=B^{(k)}$ with probability $1-\min \{1,R\}$
   \EndFor\label{euclidendwhile}
   \State \textbf{return} All the $B^{(k)}$'s and $\theta^{(k)}$'s.
\end{algorithmic}
\end{algorithm}

\subsection{Split-merge Metropolis-Hastings algorithm}
In order to improve the Metropolis-Hastings sampling efficiency on partition
$B$ in terms of number of blocks or clusters,
we propose a split-merge algorithm.
The details of splitting and merging operations and calculations
of the transition probabilities
$q(B^*|B^{(k)})$ and $q(B^{(k)}|B^*)$ are described as follows.
We assign the probabilities $(p_s,p_m,p_k)$
for splitting a cluster,
merging two clusters,
or keeping the previous partition.
For example, $(0.475,0.475,0.05)$ is the default setting in our experiments.

For the splitting action, a cluster is randomly selected
with a probability proportional to
its within-cluster distance.
Here we consider two distances:
(1) the average of all
pairwise distances between
two observations,
$$
\frac{1}{n_{b} (n_{b}-1)}\sum_{i,j\in b}\|Y_i-Y_j\|_2,
$$
where
$n_{b}$ is the number of elements in a cluster $b$
and $\|\cdot\|_2$ is the Euclidean norm
(note that it does not need to specify $i\neq j$ since $\|Y_i-Y_j\|_2=0$ when $i=j$), and
(2) the maximum of all
pairwise distances in cluster $b$
$$\max_{i,j\in b} \|Y_i-Y_j\|_2.$$

After the cluster is selected, if there are only two points in the cluster,
then it is separated into two clusters directly.
Otherwise, we find two observations with the largest pairwise distance,
and use them as the cores of the new two clusters,
and then independently assign the rest points
with the probability proportional to their distances with these two cores.
Furthermore, we allow one core to
jump to the other cluster with a small probability, say 0.01.
The probability of the resulting partition by splitting is the
product of choosing a cluster, the points assigning to the cores, and
the jumping the core, say $prob^*$.
Therefore, the transition probability $q(B^*|B^{(k)})=p_s\times prob^*$
 (recall that $p_s$ is the splitting probability).
By doing this we have a positive backward transition probability in all possible cases
to guarantee the aperiodicity of the Markov chain.

For the merging action, there are four options of between-cluster distances:
(1) the average of all the pairwise distances crossing the two clusters
$b_1$ and $b_2$
$$
\frac{1}{n_{b_1} n_{b_2}}
\sum_{i\in b_1, j\in b_2} \|Y_i-Y_j\|_2,
$$
(2) the maximum of pairwise distances crossing the two clusters
$$
\max_{i\in b_1,j \in b_2} \|Y_i-Y_j\|_2,
$$
(3) the minimum of pairwise distances from two clusters
$$
\min_{i\in b_1,j\in b_2} \|Y_i-Y_j\|_2,
$$ and
(4) the Hausdorff distance between the two clusters
$$
\max\{\max_{i \in b_1}\min_{j\in b_2} \|Y_i-Y_j\|_2,\max_{j \in b_2}\min_{i \in b_1} \|Y_i-Y_j\|_2\}.
$$

A pair of clusters is sampled with
the probability that is proportional to the reciprocal
of their between-cluster distance, say $prob^*$.
Therefore, the transition probability $q(B^*|B^{(k)})=p_m\times prob^*$
 (recall that $p_m$ is the merging probability).
 The backward transition probability is one of the following three cases.
Case 1: If the two merged clusters can be obtained by
the splitting action without jumping a core,
then the backward transition probability is
the product of the splitting probability,
$p_s$, the probability of selecting the two cores,
and the probability of assigning the rest samples
to the selected cores.
Case 2: If the two merged clusters can be obtained
 by the splitting action with jumping a core to
 the other cluster, then the backward transition probability
 is the one in case 1 multiplied by the
 jumping probability.
Case 3: If the two merged clusters cannot be obtained
 by either cases 1 or 2, then the backward transition probability is zero.

\begin{algorithm}[H]
\caption{Split-Merge MCMC algorithm}\label{Gibbs2}
\begin{algorithmic}[1]
   \State Update $B$ and $\theta$.
   \For{$k =1:N$}\Comment{$N$ is the number of total iterations. Suppose that the current values are $\theta^{(k)}$ and $B^{(k)}$.}
      \State Randomly sample $\theta^{(k+1)}$ from the discrete posterior of $\theta$
      \State Randomly choose splitting, merging, or remaining the same with probabilities $(p_s,p_m,p_k)$.
      \State Do splitting or merging actions as described in the previous paragraph, call the new partition $B^*$.
      \State Calculate $$R = \frac{p_n(B^*|n,\lambda)L_p(B^*| \theta^{(k+1)},{\cal G} Y)q(B^*|B^{(k)})}{p_n(B^{(k)}|n,\lambda)L_p(B^{(k)}|\theta^{(k+1)},{\cal G} Y)q(B^{(k)}|B^*)},$$
       where the transition probabilities $q(B^*|B^{(k)})$ and $q(B^{(k)}|B^*)$ depend on splitting or merging action applied.
      \State Accept $B^{(k+1)}=B^{*}$ with probability $\min \{1,R\}$
      \State Keep $B^{(k+1)}=B^{(k)}$ with probability $1-\min \{1,R\}$
   \EndFor\label{euclidendwhile}
   \State \textbf{return} all the $B^{(k)}$'s and $\theta^{(k)}$'s.
\end{algorithmic}
\end{algorithm}

It is important to show that our proposed split-merge MCMC algorithm converges
 to its stationary distribution regardless of the initial state.
Since we leave a small probability that the partition keeps the same in the Gibbs sampling
and the discrete posterior of $\theta$ stays positive always,
then the transition probability
$$
p_n(\theta^{(k+1)},B^{(k+1)}|\theta^{(k)},B^{(k)})>0,
$$
where $\theta^{(k+1)}=\theta^{(k)}$ and $B^{(k+1)}=B^{(k)},$ and then the $(\theta,B)$-valued Markov chain constructed by Algorithm 2 is aperiodic.
\begin{lem}
The $(\theta,B)$-valued Markov chain constructed by Algorithm 2 is aperiodic.
\end{lem}

Since there is positive chance that the partition can be split
further into a simplest partition in which each element is a cluster,
then all possible partitions
communicate with each other,
 so that the $(\theta,B)$-valued Markov chain constructed by Algorithm 2 is irreducible.
 Given that the sample size $n$,
the size of the state space of $B$ known as the Bell number \citep{Bell1934},
 and the size of the state space of $\theta$
are all finite,
then the irreducibility also implies positive recurrence.
Consequently, the $(\theta,B)$-valued Markov chain constructed by Algorithm 2 is ergodic \citep{ISaMad1976}.

\begin{lem}
The $(\theta,B)$-valued Markov chain constructed by Algorithm 2 is irreducible, and thus is positive recurrent.
\end{lem}

\begin{thm}
(Ergodic theorem) The $(\theta,B)$-valued Markov chain constructed by Algorithm 2 converges to its stationary distribution
$p_n(\theta,B|{\cal G}Y)\propto p(\theta)\times p_n(B|n,\lambda) \times L_p(\Gamma^{-1}|{\cal G}Y)$.

\end{thm}

\section{Experiments}
We test the proposed Baysian cluster process with
Algorithm 2 on both synthetic and real data.
The initial partition $B$ is set as
$I_n$ in which each observation is a block, and
target the expected partition or
the estimated similarity matrix
$$
S=\sum_{k=n_0+1}^N\frac{ B^{(k)}}{N-n_0},
$$
where $n_0$ is the number of burn-in iterations.
Furthermore, we also define a distance matrix $D$
as $\bm{ 1}_n\bm{ 1}_n^\intercal-S.$
The distance matrix, $D$, can be expressed by a heatmap
which represents a matrix with grayscale colors with white as $1$, black as $0$, and the
spectrum of gray as values between $0$ and $1.$
Additionally, $D$ can be used as
the distance for the distance-based dendrogram \citep{Everitt1998}
to represent the hierarchical relationships of the samples.
Here we apply the single-linkage tree in our experiments \citep{Gower1969, Sibson1973}.
\subsection{Synthetic data}

\noindent {\bf Four clusters on the vertices of a unit square data}

\begin{figure}[!ht]
  \centering
 \includegraphics[width = 1.15\textwidth]{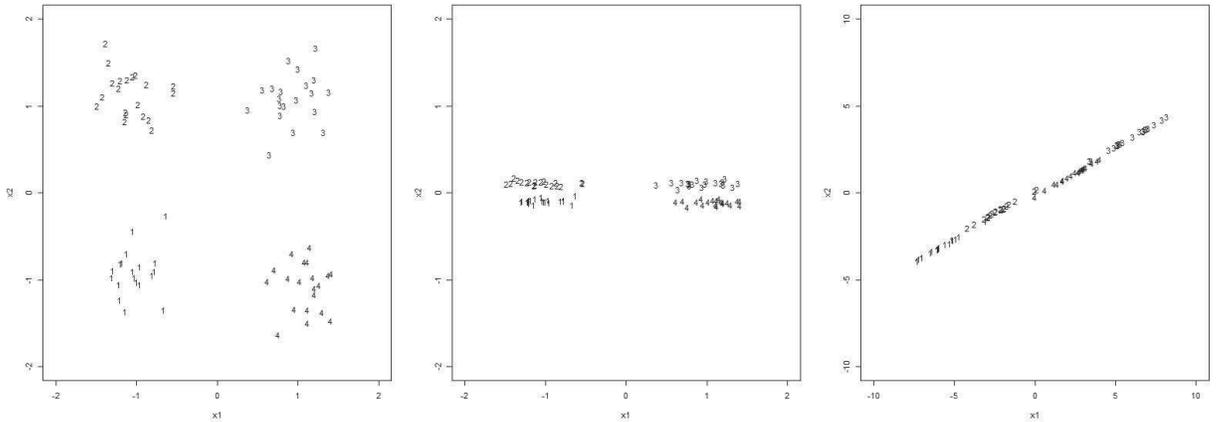}
    \caption{The scatter plots for $X_I$, $X_{II}$, and $X_{III}$
     of the unit square synthetic data from the left to the right.
    The most left panel is the original features which have four clusters at
    the vertices of the unit square with equal size $20$;
    the middle panel is the features which are transformed by
    scaling each dimension differently,
    clusters 1 and 2 are grouped as well as clusters 3 and 4 are grouped;
     the right panel shows the transformed features are aligned as a straight line.}
    \label{fig1}
  \end{figure}

We first applied the proposed cluster process with model I
on the synthetic data for four clusters centered at the four vertices of a unit square.
For each vertex $\mu_k$, we generate 20 points from
$N(\mu_k,(1/4)I_2)$ for $k=1,\ldots,4$ (see Figure~\ref{fig1}, the left panel).
We call the data $X_I$, and then apply model I to cluster $X_I$
with the average within- and between-cluster distances.
With $500$ burn-ins we use the 1000 Markov chains of $B$ samples to
calculate $D$.
The resulting heatmap and tree both successfully reveal the true clusters for most of the points (not shown here).

Then we transform the data by $X_{II}=X_I\times \left(
                                                \begin{array}{cc}
                                                  3 & 0 \\
                                                  0 & 1/3 \\
                                                \end{array}
                                              \right)
.$
The transformed features seem to have two groups (see Figure~\ref{fig1}, the middle panel),
clusters $(1,2)$ and clusters $(3, 4)$.
The cluster process with model I does not work well
for this case, while the heatmap and tree
based on model II without knowing the transformation do correctly reveal the true clusters for most of the points
with we use the $2000$ iterations after $500$ burn-in iterations (not shown here).

\begin{figure}[!ht]
  \centering
 \includegraphics[width = 0.95\textwidth]{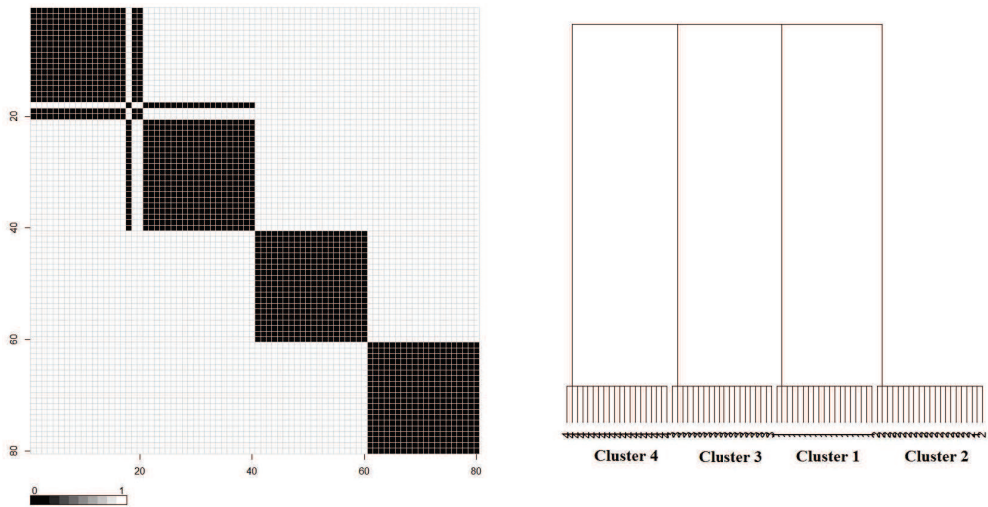}
    \caption{The heatmap of the distance matrix
    and tree both successfully reveal the true clusters for most
    of the transformed data $X_{III}$.}
    \label{fig2}
  \end{figure}

Furthermore we transform the data by $X_{III}=X_I\times \left(
                                                \begin{array}{cc}
                                                  4.1 & 2.1 \\
                                                  1.9 & 1.1 \\
                                                \end{array}
                                              \right)
.$ The transformed features are aligned in a straight line (see Figure~\ref{fig1}, the right panel).
The transformed data $X_{III}$ is more difficult to
cluster than $X_I$ and $X_{II}$, since the original four clusters are transformed to be not well separated.
After $500$ burn-in iterations, we use the $2000$ Markov chains of $B$ samples based on model III to
calculate $D$.
The resulting heatmap and tree both correctly reveal the true clusters for most of the points (see Figure~\ref{fig2}).

\noindent {\bf Two half-moons data}

\begin{figure}[!ht]
  \centering
 \includegraphics[width = 0.95\textwidth]{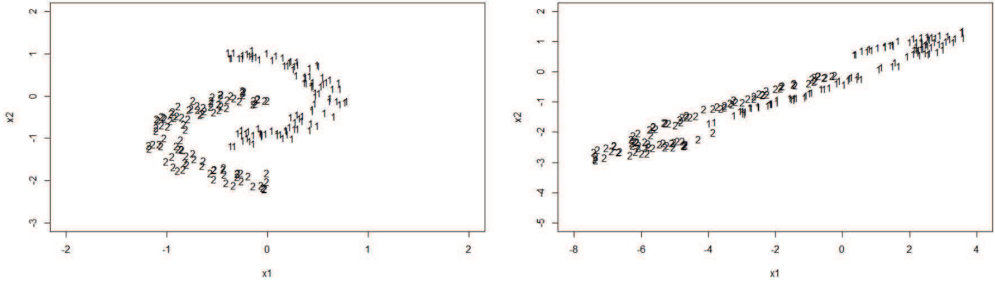}
    \caption{The scatter plots of the original and transformed two half-moons data.
    Both of the original and the transformed data are difficult for clustering since the clusters are not convex.
    }
    \label{fig3}
  \end{figure}

Further, we apply our approach with model II to
the famous two half-moons data (see Figure~\ref{fig3}, the left panel)
which is generated by the R package of `clusterSim'
\citep{WalDudek2012}
with the formula as follows
\begin{eqnarray*}
  (-0.4+|r\times \cos (\alpha)|,r\times \sin (\alpha)) &\mbox{for the first half-moon shape} \\
  (-|r\times \cos (\alpha)|,r\times \sin (\alpha)-1) &\mbox{for the second half-moon shape},
\end{eqnarray*}
where
$
r\sim \mbox{Uniform}(0.8,1.2),
$
and
$
\alpha \sim \mbox{Uniform}(0,2\pi).
$
Both two clusters are not convex.
Consequently, it makes distance-based clustering methods
such as $K$-means and distance-based hierarchical clustering
\citep{Everitt1998, Jain1999} even more difficult
to identify the correct clusters.
We use the average between-cluster distance
and the minimum within-cluster with $1000$ iterations after $610$ burn-in iterations.
The resulting heatmap and model tree both show the two half moons clearly.

In contrast, due to non-convex clusters,
the classical $K$-means centering with $K=2$ cannot correctly identify
the two half-moons clusters by assigning
two convex clusters with centers $(0.1521,0.0922)$ and
$(-0.5762,-1.2349)$, respectively.
The error rate of the $K$-means approach with $K=2$ is $0.23$
while the average error rate of our approach with model II is $0.115$.

\begin{figure}[!ht]
  \centering
 \includegraphics[width = 0.95\textwidth]{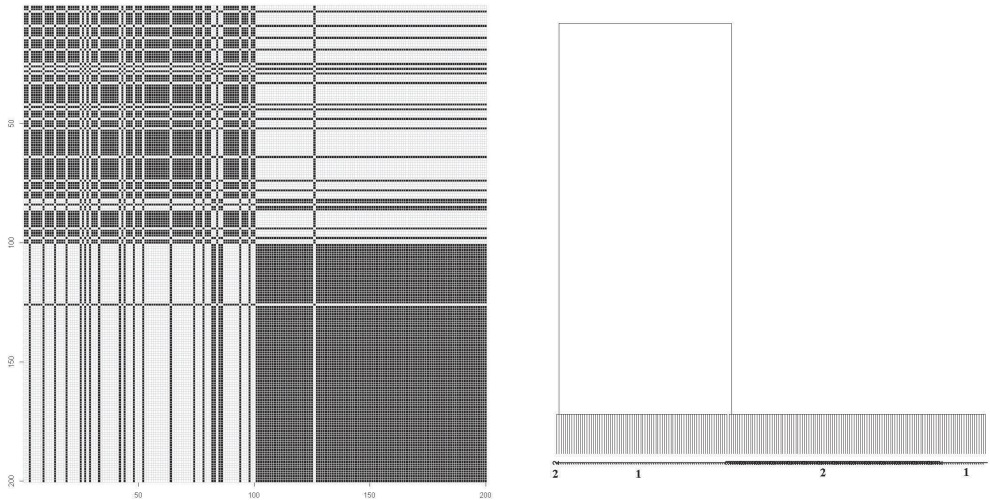}
    \caption{The heatmap of the distance matrix and single-linkage tree both successfully reveal the true clusters
    for most of the transformed two half-moons data.}
    \label{fig4}
  \end{figure}

We further transformed the two half-moons data by
$$
Y' = YA, \; \mbox{where }
A= \left(
      \begin{array}{cc}
      4.1 & 1.1 \\
      2.1 & 1.1 \\
     \end{array}
     \right),
     $$
and apply our approach based on model III with the maximum within-cluster distance
and the minimum between-cluster distance.
After the transformation, the two half-moons clusters become thin and long, and are still
not well separated.
However, our approach with model III can still recover the clusters successfully according to
the resulting heatmap and tree (Figure~\ref{fig4}) with $1000$ iterations after $400$ burn-in iterations.
The error rate of the $K$-means approach with $K=2$ is $0.15$
while the average error rate of our approach with model III is $0.11$.

\subsection{Real data}

\noindent {\bf Model I: Gene expression data of Leukemia patients}

\begin{figure}[!ht]
  \centering
 \includegraphics[width = 0.55\textwidth]{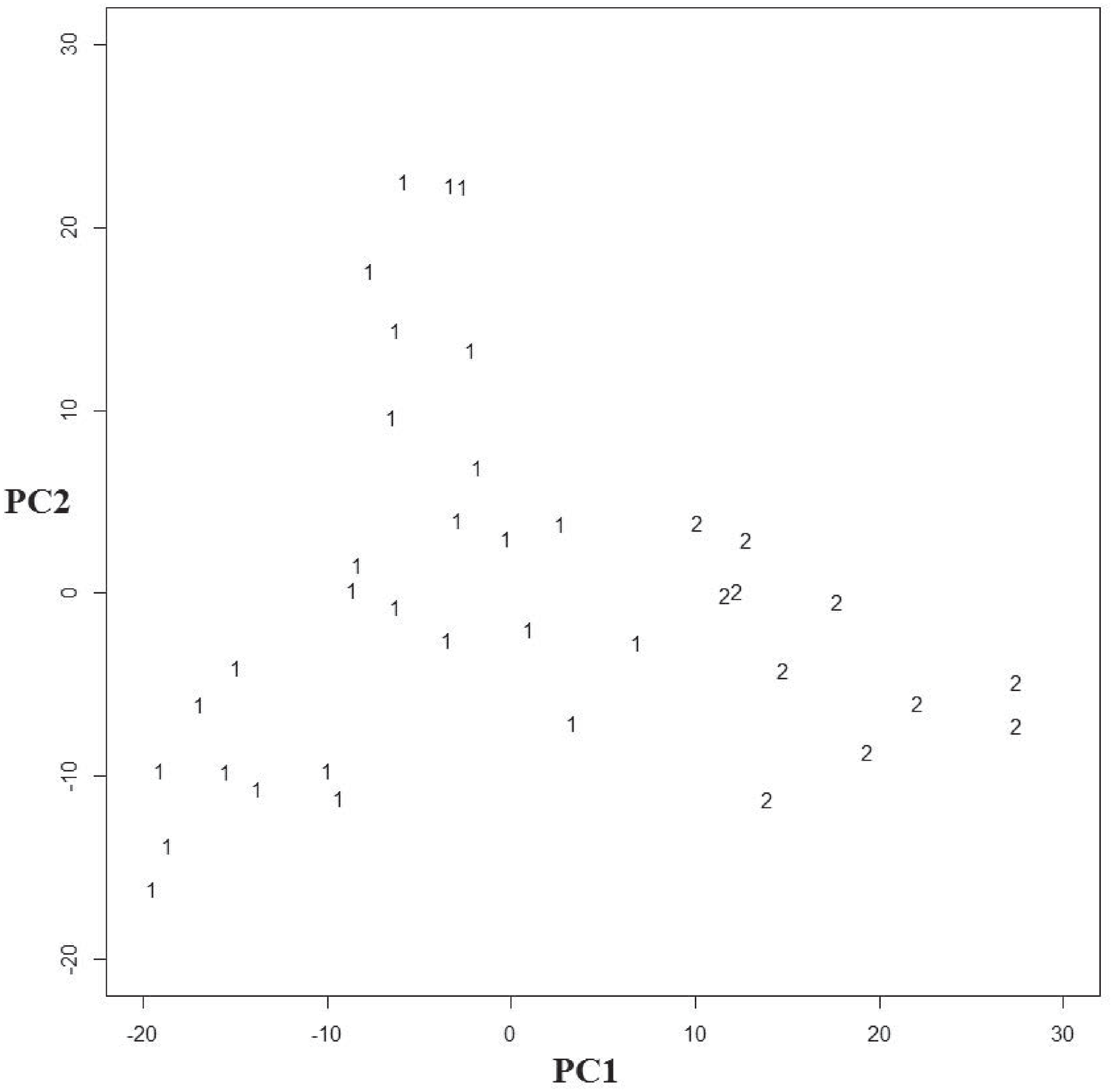}
    \caption{The scatter plot of the leukemia data after projecting the original data
    onto the subspace spanned by the first two principal components
    $(PC_1,PC_2)$.}
    \label{fig5}
  \end{figure}

\begin{figure}[!ht]
  \centering
     \includegraphics[width = 0.95\textwidth]{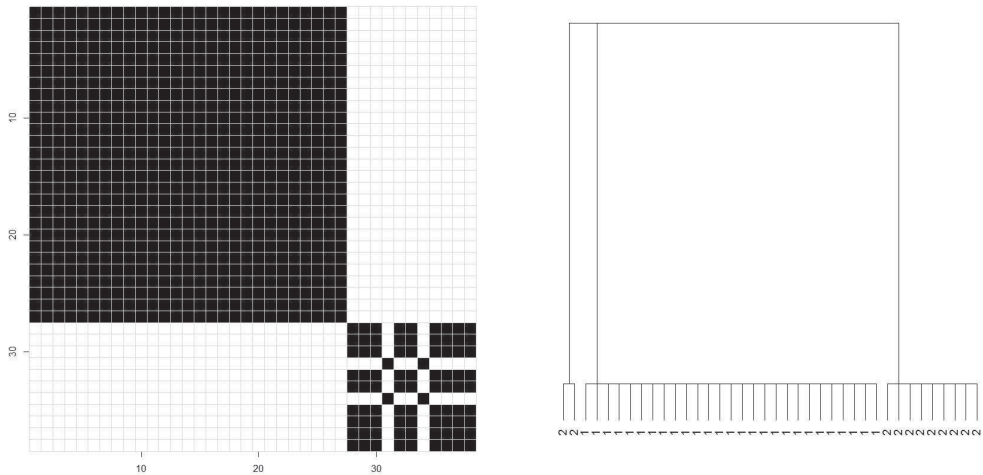}
  \caption{Both the heatmap of the distance matrix and single-linkage tree
  reveal the two clusters of the leukemia data for most of the points.}
  \label{fig6}
\end{figure}

Besides the synthetic data, we also evaluate the performance of the proposed
approach by using real data.
The gene expression microarray data \citep{Lichman2013}
has been used to study genetic disorder such as
identifying  diagnostic or prognostic biomarkers or clustering and classifying diseases \citep{Dudoit2002}.
For example, \cite{Golub1999} classified patients of acute leukemia into two
sub types, Acute Lymphoblastic Leukemia (ALL) and Acute Myeloid
Leukemia (AML).
For illustration purpose, we use the training set of the leukemia data which consists of 3051 genes and 38 tumor mRNA samples.
Pretending we do not know the label information, we would like to cluster the $38$ samples according to
their $3051$ features (gene expression levels).
The two clusters comprise $27$ ALL cases and 11 AML cases.
Since the number of features is larger than the sample size, our approach is not applicable to this dataset directly.
Therefore, we first reduce the dimension by projecting the data on the subspace
which consists of the first twenty principal components (PC) \citep{Jolliffe1986}.
Note that these PCs are orthonormal which satisfies the assumption of model I.
We show the scatter plot of the leukemia data
after projecting the original data onto the subspace spanned by the first two principal components
    $(PC_1,PC_2)$ in Figure~\ref{fig5}.
The resulting tree and heatmap based on model I (Figure~\ref{fig6})
both reveal the true clusters with the average
within- and between-cluster distances and $1000$ iterations after $500$ burn-in iterations.

\noindent {\bf Model II: Wine data}

\begin{figure}[!ht]
  \centering
 \includegraphics[width = 0.95\textwidth]{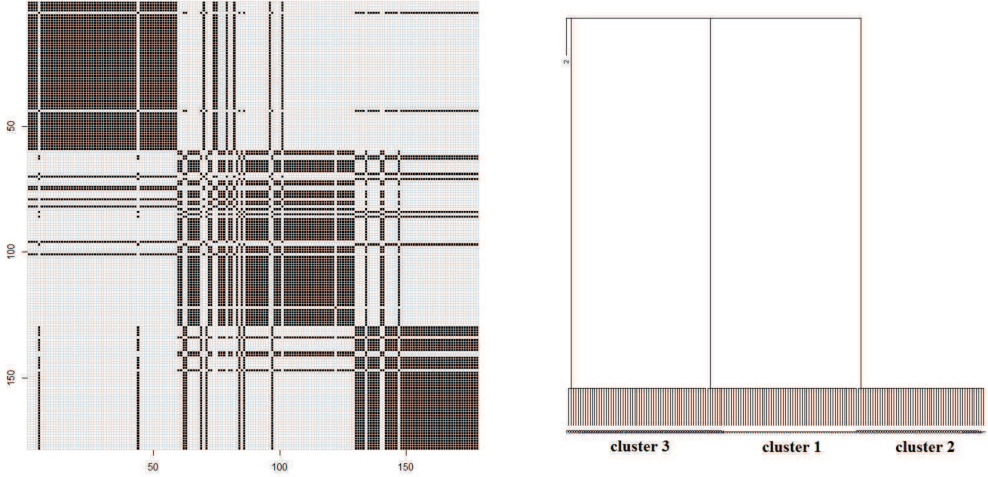}
   \caption{Both the heatmap of the distance matrix and single-linkage tree reveal the three clusters of the wine data for most of the points.}
   \label{fig7}
  \end{figure}

We also explore the benchmark wine data \citep{Lichman2013}.
These data are the results of a chemical analysis of wines grown in the same region in Italy but derived from three different cultivars clusters.
The cluster sizes are 59, 71, and 48, respectively.
The clustering analysis is based on
 the 13 attributes of the three types of wines with 178 observations.
Based on model I, we run $1500$ iterations after $1350$ burn-in iterations with
the average between-cluster distance, the minimum
within-cluster distance, and
$(p_s,p_m,p_k)=(0.09,0.90,0.01)$.
The heatmap and tree (Figure~\ref{fig7}) both show that
the proposed approach with model I can identify the tree clusters
for most of the points.

\noindent {\bf Model III: Geographic coordinate system data of Denmark's 3D Road Network}

\begin{figure}[!ht]
  \centering
 \includegraphics[width = 0.95\textwidth]{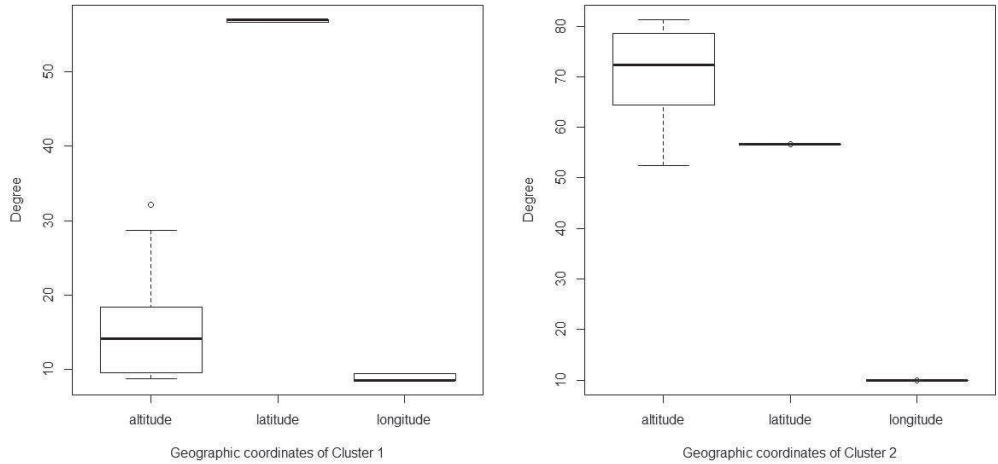}
   \caption{The boxplots show that the altitude has high higher variances for both clusters 1 and 2.
   The boxplots also indicate that both clusters have higher variances of altitude than latitude and longitude, and
   the two clusters have very different altitude and latitude degrees on average.}
   \label{fig8}
  \end{figure}

\begin{figure}[!ht]
  \centering
 \includegraphics[width = 0.95\textwidth]{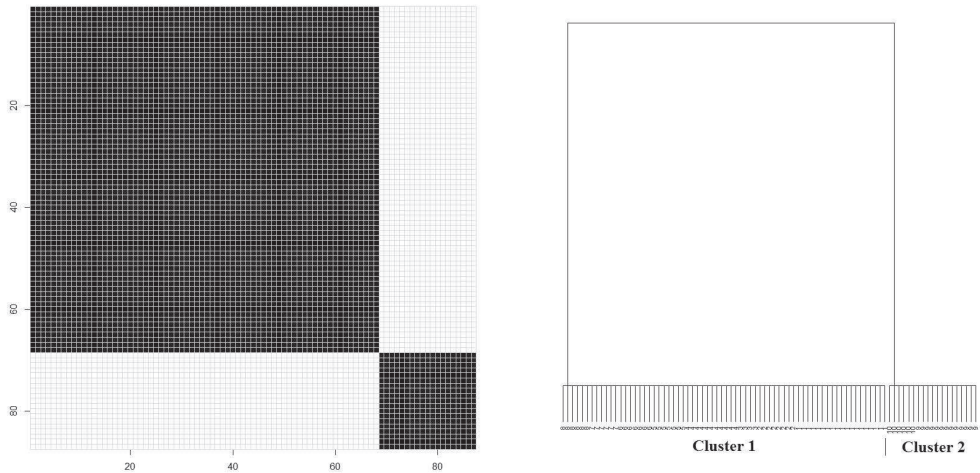}
   \caption{Both the heapmat of the distance matrix and single-linkage tree reveal the two true clusters of the Denmark 3-D road map data.}
   \label{fig9}
  \end{figure}

This three dimensional road network dataset of
geographic coordinates includes the altitude, latitude, and longitude
degrees of each road segments in North Jutland in northern Denmark,
which is publicly available at
the UC Irvine Machine Learning Repository \citep{Kaul2013, Lichman2013}.
We obtain 87 observations of 10 different objects
which belong to two clusters (cluster 1: objects 1 to 8; cluster 2: objects 9 to 10) based on
their longitude and latitude degrees.
Note that each objects may have several observations measured from different angles,
and the altitude values are extracted from NASA's Shuttle Radar Topography Mission (SRTM) data \citep{Jarvis2008}.
The number of observations of the objects varies from four to twenty.
The average geographic coordinates of cluster 1 are
$(14.9137,56.9522,8.7564)$, which are the altitude, latitude, and longitude degrees, respectively,
and the average geographic coordinates of cluster 2 are
$(70.2441,56.6335,9.9938).$
The standard deviations of the altitude, latitude, and longitude degrees of cluster 1 are
$(5.2903, 0.1330, 0.3750)$, and the standard deviations of cluster 2 are
$(9.6316, 0.0013, 0.0054).$
The longitude and latitude degrees determine the true clusters, and they both have much smaller variances than the altitude
(see the boxplots in Figure~\ref{fig8}).
The resulting heatmap and tree (Figure~\ref{fig9}) both show the true two clusters with the average
within- and between-cluster distances and $1000$ iterations after $500$ burn-in iterations.

\section{Discussion}
We have presented a Bayesian clustering approach
 which is invariant to
 different groups of affine transformations.
These problems are dealt with an exchangeable partition
prior which avoids label-switching problems and the
profile likelihoods under three types of covariance structures.
Note that the proposed approach
does not target the partition
maximizing the posterior distribution.
Instead, it estimates the expected partition
or the distance matrix, which is
more reliable for a moderate sample size.
It works reasonably well
across various applications.
Additionally, the transition probability ratio is influenced by
the choice of between- and within-cluster distances and the split and merge probabilities $(p_s,p_m)$.
For example, we choose the minimum
within-cluster distance and $(p_s,p_m,p_k)=(0.019,0.98,0.001)$ for both
the original and transformed two half-moons data.
However, we apply
the average between-cluster distance for the original data, but
the maximum between-cluster distance for the transformed data.
Moreover, when applying other types of the proposed within-cluster distances,
the proposed split-merge algorithm does generate the desired clusters
after $2000$ burn-in iterations in our experiments.
The minimum between-cluster distance tends to connect two nearest clusters and
produce a long cluster where neighboring elements
in the same cluster have small distances.
remain a cluster with a small with-cluster distance \citep{Gower1969, Sibson1973}.
Therefore, we obtain a posterior mean partition matrix instead of a maximum
likelihood estimate of partition.

The main contributions of our work include:
1) The proposed three clustering models with
three types of covariance structures can handle general cases
of affine transformations. In contrast, \cite{Vogt2010} only
dealt with the case of model I.
2) The split-merge algorithm can generate partition candidates for the
Gibbs sampling
much more efficiently (not shown here) than
the classical Algorithm 1.
It also ensures that the resulting partition-valued
Markov chain is ergodic and convergent in distribution.
3) The experiments show the advantages of our cluster
process which successfully identifies
 the true clusters using the proposed distance matrix.
In particular if the clusters are not well separated, the distance matrix with
probabilistic nature can still reveal the relationships through
hierarchical approaches.

The proposed method could be used to extract interesting information from
aerial photography, genomic data, and data with attributes under different scales,
especially when the nearest neighbors may belong to different clusters in the feature space.
R code for implementing the proposed clustering method can be obtained upon request.

% ** Acknowledgements **
\section*{acknowledgement}
The authors thank Professor Peter McCullagh for his insightful comments and suggestions on an early version of this paper which substantially improved the quality of the manuscript.
This research is in part supported by the LAS Award for Faculty of Science at the University of Illinois at Chicago and the
In-House Award at the University of Central Florida.

% ** The bibliograhy **
%\bibliographystyle{ba}
%\bibliography{<bib-data-file>}% place <bib-data-file>

\end{document}